\documentclass[%
 reprint,
superscriptaddress,
 amsmath,amssymb,
aps,
 physrev,
 showkeys,
longbibliography]{revtex4-2}

\usepackage{graphicx}
\usepackage{dcolumn}
\usepackage{bm}
\usepackage{subcaption,amsmath,gensymb,esvect,xcolor,xr,multirow,commath}
\usepackage[justification=raggedleft,singlelinecheck=false]{caption}
\usepackage[colorlinks=true,citecolor=blue]{hyperref}
\usepackage[normalem]{ulem}

\newcommand{\etal}{\textit{et al.}}
\newcommand{\mycomment}[1]{}

\usepackage{etoolbox} 
\usepackage{lipsum} 
\usepackage[capitalize]{cleveref}
\usepackage{xr-hyper} 
\makeatletter
\appto{\appendix}{%
  \@ifstar{\def\theequation@prefix{A.}}%
          {}%
}
\makeatother

\begin{document}

\preprint{APS/123-QED}
\title{Magneto-optical Response of 5-SL MnBi$_2$Te$_4$ in Spin-Flip States}

\author{Shahid Sattar}
 \email{shahid.sattar@lnu.se}
\affiliation{Department of Mathematics and Physics, Linnaeus University, SE-39231 Kalmar, Sweden}
\author{Roman Stepanov}
\affiliation{Department of Mathematics and Physics, Linnaeus University, SE-39231 Kalmar, Sweden}
\author{A. H. MacDonald}
\affiliation{Department of Physics, University of Texas at Austin, Texas 78712, USA}
 \author{C. M. Canali}
\affiliation{Department of Mathematics and Physics, Linnaeus University, SE-39231 Kalmar, Sweden}
 \date{\today}

\date{\today}

\begin{abstract} 

Magneto-optical effects like Kerr and Faraday rotations provide a direct probe of topological order in thin films of the 
magnetic topological insulator MnBi$_2$Te$_4$ (MBT). Motivated by recent experimental studies of spin-flip/flop transitions in MBT thin films, we investigate the interplay between interlayer spin configurations, topological order, and magneto-optical response in five 
septuple-layer (5-SL) MBT using first-principles calculations and a simplified coupled-Dirac-cone model. 
Our results reveal that, despite possessing a non-zero out-of-plane magnetization, 5-SL MBT thin films can be either 
${\cal C}=+1$ topological insulators or ${\cal C}=0$ topologically trivial insulators 
depending on the relative spin orientations of the top and bottom SLs. We evaluate the Faraday and Kerr rotation angles using
tight-binding models derived from \textit{ab-initio} calculations and by comparing our results with those of a simplified 
coupled Dirac-cone model clarify the macroscopic mechanisms underlying the magneto-optical response of spin-flip states. 
These theoretical findings highlight the tunability of topological and magneto-optical properties in MBT thin films
and provide microscopic insight into the emergence of complex topological order in layered antiferromagnetic materials.

\end{abstract}

\maketitle

\par 
\section{Introduction}
Time-reversal symmetry breaking can be induced in topological insulators by external magnetic fields \cite{valdes2012terahertz,wu2016quantized}, magnetic doping \cite{okada2016terahertz,mogi2022experimental} or proximity effects \cite{lang2014proximity,che2018proximity} and 
detected by measuring their Faraday and Kerr rotation magneto-optical responses. 
In intrinsic magnetic TIs such as MnBi$_2$Te$_4$ (MBT), magnetization survives down to the few-layer limit in thin films 
and couples to the topological surface states \cite{zhang2019topological,li2019intrinsic,otrokov2019unique}. 
In their ground state MBT thin films, composed of van der Waals (vdW) gapped and antiferromagnetically (AFM) coupled 
neighboring septuple layers (SLs), are Chern 
insulators exhibiting the quantum anomalous Hall effect (QAHE) for odd SLs \cite{deng2020quantum,liu2020robust}, while for even SLs they 
realize the axion insulating phase expected to display the topological magnetoelectric effect \cite{liu2020robust,liu2021magnetic,lin2022direct}. Other distinctive even–odd-layer-dependent quantum effects have also been actively investigated 
\cite{zhao2021even,ovchinnikov2021intertwined,lupke2022local,li2024fabrication,chen2024even}.

While transport measurements are commonly used to identify these topological phases \cite{li2024fabrication}, 
magneto-optical response provide a powerful complementary tool for topological characterization of MBT thin films. 
For example, Qiu \etal\, have confirmed the existence of dynamical axion quasiparticles in 6-SL MBT using ultrafast optical pump–probe experiments \cite{qiu2025observation}. Quantized magneto-terahertz effects (Faraday and Kerr rotations) were reported in 6-SL and 7-SL MBT thin films in Ref. \cite{han2025quantized} and helicity-dependent optical control of AFM order has been demonstrated
in 6-SL MBT films \cite{qiu2023axion}.  Real-time observation of magnetization and magnon dynamics across different thicknesses
are opening new routes for ultrafast manipulation of topological and magnetic order in MBT samples \cite{bartram2023real}. 

While A-type AFM order with out-of-plane spins is the ground state in MBT \cite{zhang2019topological,otrokov2019unique}, 
under an applied external magnetic field thin films exhibit rich and complex spin configurations. The MBT 
phase diagram includes field-induced spin-flip/flop transitions that modify the spin-wave excitations (magnons) of the AFM-ordered SLs \cite{bartram2023real,sass2020robust,qian2023spin}. For example, Lian \etal\, have studied the QAHE under spin flips/flops in 7-SL MBT by tuning the gate voltage and the perpendicular magnetic field and have uncovered a cascade of quantum phase transitions \cite{lian2025antiferromagnetic}. Experimental work by Han \etal\,discussed abrupt magnitude changes in magneto-optical response during these transitions, suggesting the possible existence of a Chern insulator state even in 6-SL MBT films under external magnetic field \cite{han2025quantized}. 

Existing theoretical studies of TIs \cite{tse2011magneto} and magnetic TIs \cite{lei2023kerr}, however, mostly focus on the ground state and discuss layer-dependent Faraday and Kerr rotations and their electric field dependence. 
It is unclear how the topological order, relative energies and bandgaps change in moving from the AFM ground state (alternating spin-alignment in the out-of-plane direction) to spin-flip states, and eventually to the ferromagnetic (FM) spin-alignment state. 

Here, we focus specifically on these alternative microscopic spin flip configurations of 5-SL MBT thin films, and map out the corresponding energy landscape. The interplay between different interlayer spin-arrangements, bandgaps, and frequency-dependent transverse and longitudinal conductance and magneto-optical responses are discussed using first-principles calculations and compared with a 
simplified coupled-Dirac-cone toy model \cite{lei2023kerr}. We argue that despite possessing a non-zero out-of-plane magnetization, a 5-SL thin film can exhibit either non-vanishing (${\cal C}=+1$) or vanishing (${\cal C}=0$) Chern insulating phases.  The Chern number of the 
ground state instead depends mainly on the relative spin orientations of the top and bottom SLs.  We 
clarify the merits and the limitations of the coupled-Dirac-cone model \cite{lei2023kerr} in describing the 
salient transport and magneto-optical features of these magnetic TIs.

\par 
\begin{figure}
    \includegraphics[width=8cm]{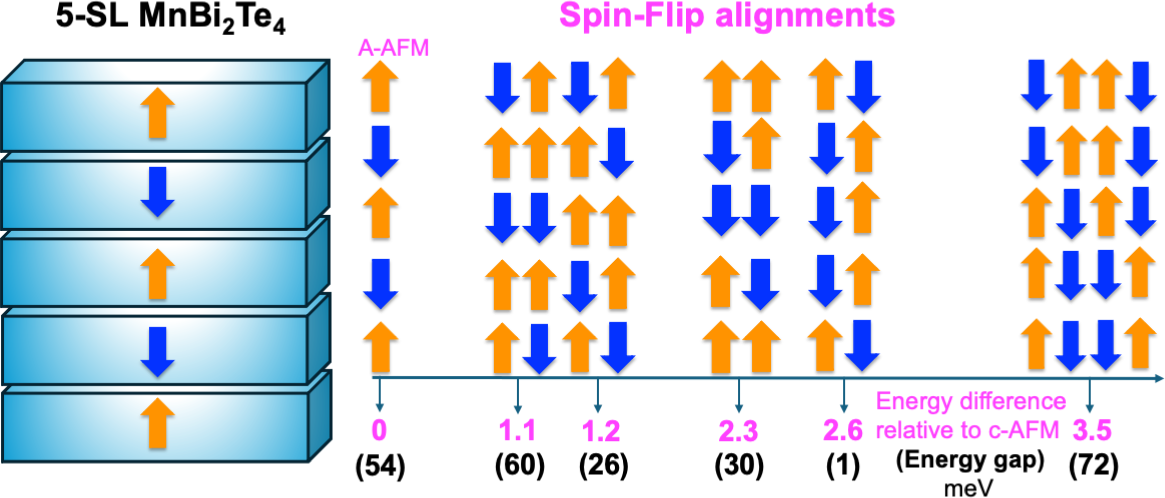}
    \caption{(left) Schematic of a 5-SL MnBi$_{2}$Te$_{4}$ in the AFM ground state configuration. (right) Alternate spin-flip 
    configurations with the same net magnetic moment of $5\mu_B$.  Energy difference per unit cell (in meV) compared to the AFM ground state and band gaps (in meV) are given at the bottom of degenerate cases in magenta and black colors, respectively.}
    \label{fig:fig1}
\end{figure}

\section{COMPUTATIONAL DETAILS}

First-principles calculations were performed using density functional theory (DFT) and the projector augmented wave method\,\cite{paw1,paw2} as implemented in the Vienna Ab-initio Simulation
Package (VASP)\,\cite{vasp}. The generalized gradient
approximation in the Perdew-Burke-Ernzerhof parametrization was employed for the
exchange-correlation potential. The plane wave cutoff energy was set to a large value of 430 eV. We
use a $\Gamma$-centered Monkhorst-Pack $9\times9\times1$ ($12\times12\times1$) k-mesh for
the Brillouin zone integration for structural relaxation (self-consistent calculations). 
In the iterative solution of Kohn-Sham equations, we ensured a
force convergence of $10^{-3}$ eV/\AA\ and an energy convergence of $10^{-7}$ eV. 
DFT-D3 van der Waals dispersion corrections were employed in all calculations to capture long range van der Waals
interactions \cite{grimme2010consistent}. A vacuum layer of 15 \AA\ was added in the
out-of-plane direction to create two-dimensional structures free from
artifacts of the three-dimensional periodic boundary conditions. 
To correctly describe strong on-site Coulomb interactions of localized Mn $d-$electrons, we used a Hubbard $U$ correction (GGA+U method) 
following Dudarev's scheme \cite{dudarev1998electron} and fixed the onsite Coulomb (U) and exchange (J) parameters 3.9 eV and 0.0 eV, respectively. 

A real-space tight binding (TB) model Hamiltonian was obtained by parameterizing the
self-consistent Kohn-Sham Hamiltonian using the Wannier90 package \cite{marzari1997maximally,mostofi2014updated,pizzi2020wannier90}. 
For this purpose, we used the VASP2WANNIER90 interface and included Mn$-d$, Bi$-p$ and Te$-p$ orbitals in generating the Wannier functions. Topological characteristics were computed using the WannierTools \cite{wu2018wanniertools} and WannierBerri packages \cite{tsirkin2021high}. 

For the computation of the optical conductivity, we used the Kubo-Greenwood formula \cite{kubo1957statistical,greenwood1958boltzmann}:
\begin{equation} 
\begin{aligned}
\sigma_{\alpha\beta}(\hbar\omega)=\frac{i e^2}{\hbar}
\int\!\frac{d^2k}{(2\pi)^2}\sum_{n m}
\frac{f_{n\mathbf k}-f_{m\mathbf k}}{E_{n\mathbf k}-E_{m\mathbf k}}\\\quad
\times\frac{\big|\langle m\mathbf k|\partial_{k_x} H_{\mathbf k}|n\mathbf k\rangle\big|\big|\langle m\mathbf k|\partial_{k_y} H_{\mathbf k}|n\mathbf k\rangle\big|}
{E_{n\mathbf k}-E_{m\mathbf k}-(\hbar\omega+i\eta)},
\end{aligned} \label{eq-1}
\end{equation}
where  $f_{n\mathbf k}$ is Fermi-Dirac distribution function giving band occupation probability, $\omega$ is the optical frequency, $\hbar$ is the
reduced Planck’s constant, $n,m$ are band indices, $|n\mathbf k\rangle$ and $|m\mathbf k\rangle$ are Bloch states and  $E_{n\mathbf k}$ and $E_{m\mathbf k}$ are the corresponding band energies, and $\eta$ is a disorder broadening parameter.

\section{Results and Discussion}

\subsection{Spin configurations in five-septuple-layer (5-SL) MBT thin films}

Bulk MnBi$_2$Te$_4$ (MBT) combines intrinsic magnetism, due to Mn magnetic moments ($\sim 5\,\mu_B$), and non-trivial band topology characterized by the topological Z$_2$ invariant \cite{li2019intrinsic}. MBT thin films consist of ferromagnetic SL (Te-Bi-Te-Mn-Te-Bi-Te) blocks 
stacked along the c axis. The ground state of a 5-SL thin film has adjacent SLs coupled antiferromagnetically, 
with spins pointing along the out-of-plane easy axis, as shown in the left panel of Fig.~\ref{fig:fig1}. For this spin configuration, the topological surface states (TSSs) display an exchange gap of 54 meV at the Dirac point. From now on, we will consider this value of the energy gap as a zero-energy reference. 

Spin configurations involving spin flips between adjacent SLs and thus deviating from the AFM ground-state configuration are possible, regardless of the mechanism responsible for their formation. Figure \ref{fig:fig1} (right) illustrates all possible spin alignments for the 5-SL film that preserve a net magnetic moment of $5\,\mu_B$. This value corresponds to the typical magnetic moment of a single Mn atom in each SL and therefore provides a natural reference for describing spin-flip states in 5-SL MBT. The relative energies of these configurations, calculated per unit cell with respect to the AFM ground state energy, are indicated below each alignment in magenta color. The higher energy spin-configurations in Fig.~\ref{fig:fig1} are all metastable and 
can be realized by cycling applied external magnetic fields and temperatures \cite{padmanabhan2022interlayer, yan2019evolution,lian2025antiferromagnetic,jie2025reconfigurable,lei2020magnetized}. To examine whether our conclusions depend on the magnitude of total magnetization, we also investigated a configuration with a higher net moment of $15\,\mu_B$, which will be discussed later in the text.

Spin-reversal processes are energetically costly but become increasingly relevant when the exchange energies between adjacent SLs are comparable to the anisotropy energy \cite{alfonsov2021magnetic} or when external perturbations, such as magnetic field, strain, or proximity effects dominate and lower the barriers separating different magnetic stacking states \cite{qi2024exchange}. 
Importantly, these deviations from the GS spin configurations have the crucial effect of modifying the 
TSS gap magnitude and even the topological symmetry of the system. 
Considerable attention has been paid to how native point and surface defects \cite{lai2021defect,garnica2022native,liu2022visualizing,tan2023distinct} and surface reconstructions \cite{sattar2025surface} affect the TSS energy gap. Our results suggest that the role of domain structures containing 
local spin flip magnetic configurations, which has been less investigated, also deserves a careful investigation. The right panel of Fig.~\ref{fig:fig1} shows all spin configurations with the same net magnetization as the ground state 
together with their relative energies and TSS energy gaps (black numbers in parentheses at the very bottom). 
We do not address the cases of canting, whose topological response has already been discussed in Ref.~\cite{bac2022topological}. 

\begin{figure}
    \centering
      \includegraphics[width=8cm]{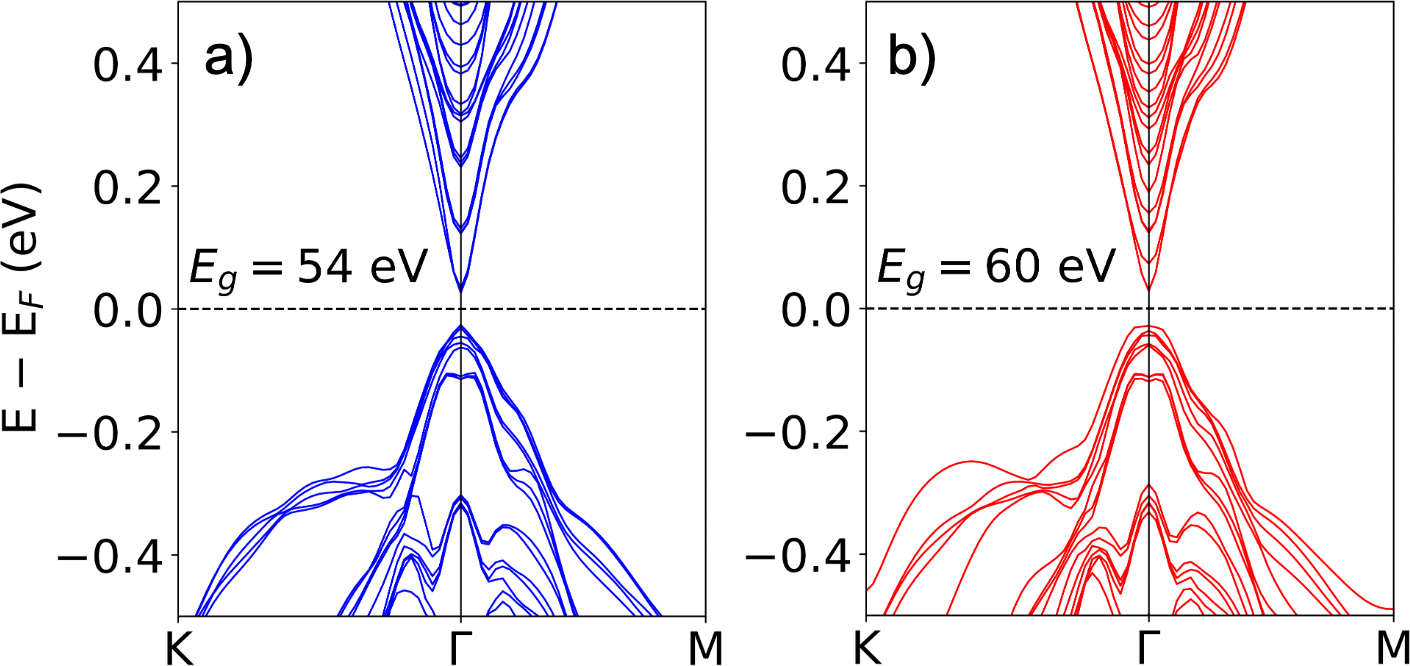}
    \caption{(a) Band structure of a 5-SL MnBi$_{2}$Te$_{4}$ in the AFM ground state (GS) configuration. (b) Band structure of degenerate spin-flip alignments having 1.1 meV higher energy than the GS.}
    \label{fig:fig2}
\end{figure}

We classify all the spin-flip configurations that preserve the same total magnetic moment of the 5-SL MBT ground state into two physically distinct categories: (1) configurations in which the spins of the top or the bottom SLs are aligned parallel to their adjacent layer(s), thereby modifying the surface magnetization; and (2) configurations in which the parallel spin alignment occurs only within the interior SLs of the 5-SL stack, leaving the surface magnetization comparatively unchanged. Within each category, we identify energetically degenerate configuration pairs, such as the first case when either top or the bottom two SLs spins are aligned in parallel (shown as orange arrows). Here, the energy relative to
the AFM ground state is only 1.1 meV per unit cell. Interestingly, we observe a flattening of the topmost valence band around the gamma point owing to this change, as shown in Fig. \ref{fig:fig2}(b). Moreover, the TSS energy gap is enhanced to 60 meV in comparison to 54 meV observed for the AFM ground state (see also Fig. \ref{fig:fig2}(a)). 
In contrast, when the SL spin alignment moves inside the film, the TSS gap decreases to 26 meV, even though
the energies of this configuration (1.2 meV) and the previous one (1.1 meV) are essentially the same. 
Hence, we conclude that the distinction between the two classes is consequential, as the TSS gap is predominantly governed by the spin alignments of the surface and adjacent interior SLs. Contiguous SLs at the surface with parallel spin 
configurations are more effective in increasing the TSS energy gap than contiguous SLs with parallel spin configurations 
located inside the film.

\par 

Extending this analysis to the last two cases (2.6 meV and 3.5 meV energy difference relative to AFM ground state) when three inner (outer) SLs are aligned parallel, we observe a TSS gap of 1 meV (72 meV), respectively. These findings are in-line with the recent experimental work of 
Ref.~\cite{lian2025antiferromagnetic}. Since the Dirac-like surface states are spatially localized largely in the top and bottom SLs only, we notice that parallel-spin SLs on the surfaces increase the local surface magnetization which, in turn, produces a larger exchange field and is responsible for an increase in the TSS gap. On the other hand, parallel spin SL  pairs in the interior reduce the effective surface exchange field thus yielding a smaller TSS gap. This behavior demonstrates that the TSS gap in MBT thin films is controlled primarily by the local top and bottom SL spin alignments rather than by the global magnetization of the sample.

\begin{figure}[!b]
    \centering
    \includegraphics[width=8.5cm]{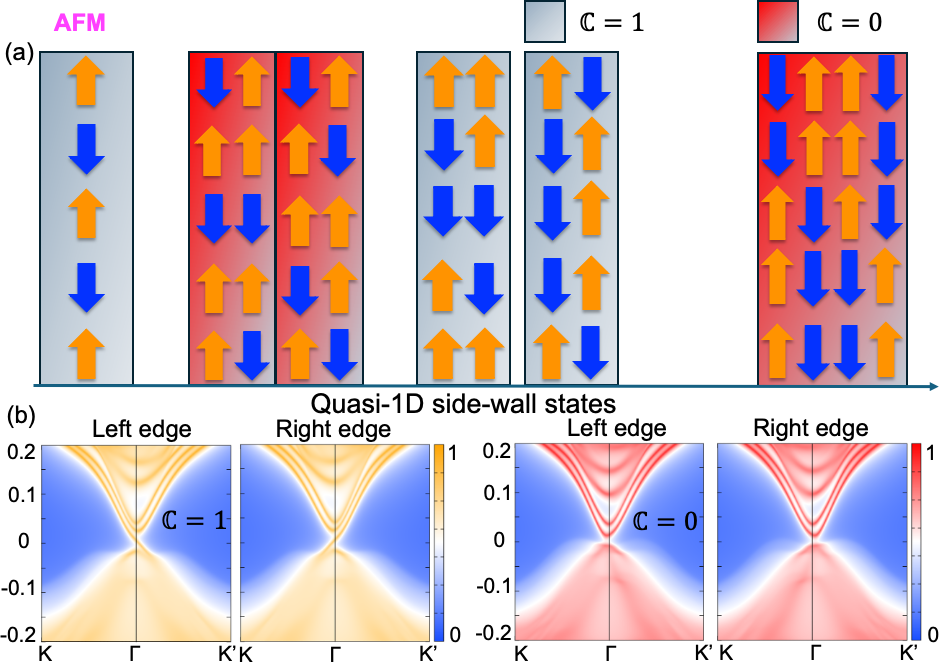}
    \caption{(a) Topological nature of different spin flip alignments. Blue columns represent ${|\cal C}|=+1$ and red columns show ${\cal C}=0$ cases, depending on the spin alignment of the top-most and bottom-most SLs. (b) Quasi-1D left and right side-wall edge states of 110 surface, shown for the two representative ${|\cal C}|=+1$ and ${\cal C}=0$ cases, respectively.}
    \label{fig:fig3}
\end{figure}

\subsection{Topological properties of MBT 5-SL thin films}
Next we examine the topological character of the spin flip states in a 5-SL MBT thin film. 
Since we have already established that the relative spin orientations of the top and bottom surface SLs play a crucial role in determining the TSS gap, it is natural to classify all cases into two groups: (1) configurations in which the surface SL spins are parallel, and (2) those in which they are anti-parallel. This distinction proves to be decisive in determining the topological nature of spin flip states by computing Chern number for each case. Interestingly, when the top-most and bottom-most SL spins are aligned in the same direction, we obtain a nontrivial Chern insulating phase with ${\cal C}=1$ as shown in blue color in Fig. \ref{fig:fig3}a. In contrast, when the surface SL spins are oppositely oriented, we find that that Chern number vanishes (${\cal C}=0$), as shown in the panels with red color in Fig.~\ref{fig:fig3}(a). In other words, the topological response mirrors our earlier conclusion that a global magnetization of the sample (5 $\mu_B$ for 5-SL MBT) does not necessarily implies a non-zero Chern number characterizing a Chern insulating state; rather, it is the relative orientation of the top and bottom SLs which dictates the topology of MBT thin films. We further confirm this by looking at a higher net magnetization of $15\,\mu_B$ as shown in Supplementary Figure. S1.

Additionally, we compute quasi one-dimensional (1D) side-wall edge states of the 110 surface for two representatives cases, with ${\cal C}=1$ and ${\cal C}=0$ respectively, as shown in Fig.~\ref{fig:fig3}(b) (left). For the ${\cal C}=1$ case, we observe spin-polarized counter-propagating chiral edge states on the left and right side-walls (orange), as expected for a quantized Chern insulating state. On the other side, the ${\cal C}=0$ case shows gapped side-wall edge states (red), which is typical of an axion insulating state. It should be mentioned that a recent experiment by Han \etal\,points to the presence of a possible Chern insulator state in an even 6-SL MBT film under an external magnetic field \cite{han2025quantized}. According to our theoretical results. this finding can be explained by a magnetic-field induced spin flip transition, where the SL spins on the top and bottom surfaces are parallel. Finally, the coexistence of ${\cal C}=0$ and ${\cal C}=1$ spin-flip states at fixed nonzero magnetization provides insight into the existence of complex topological order present in MBT thin films.

\subsection{Magneto-optical properties}

\begin{figure}[!t]
    \centering
    \includegraphics[width=8.5cm]{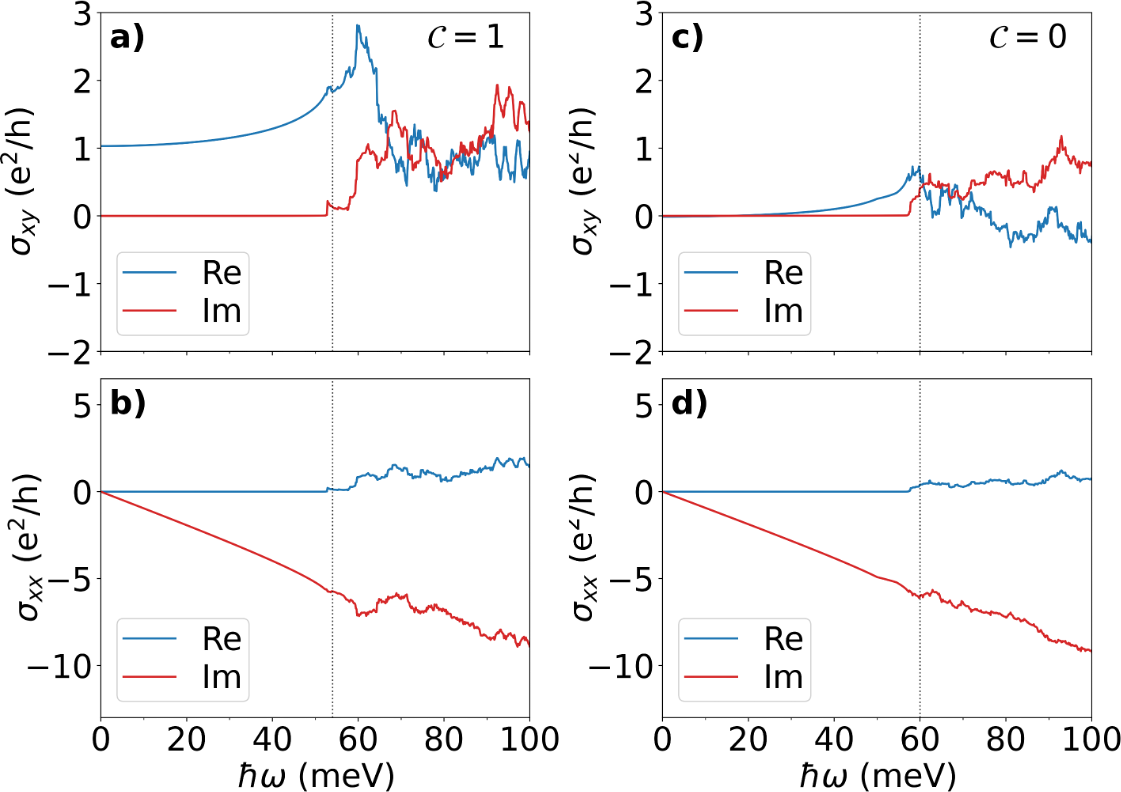}
    \caption{Frequency-dependent optical conductivity (in units of $e^2/h$) for the two representative spin-flipped 5-SL MnBi$_2$Te$_4$ cases. (a,c) transverse (Hall) conductivity ($\sigma_{xy}$) and (b,d) longitudinal conductivity ($\sigma_{xx}$) for the representative ${\cal C}=+1$ and ${\cal C}=0$. In each case, real and imaginary part is shown in blue and red colors, respectively. The TSS gap is shown as black dotted line.}
    \label{fig:fig4}
\end{figure}

Experimentally, quantum transport and magneto-optical measurements are the two most convenient ways of studying the topological properties. To elucidate the underlying mechanisms of the quantized response in characterizing a specific topological state, we first analyze the longitudinal and transverse optical conductivities. 
Using the expression of the optical conductivity tensor given in Eq.~\ref{eq-1} calculated with a TB model obtained from \textit{ab-initio} methods, we present results for the two cases (${\cal C}=0$ and ${\cal C}=1$). In Fig.~\ref{fig:fig4}, we plot the real ($\Re$, blue line) and imaginary ($\Im$, red line) parts of the transverse (a,c) and  longitudinal (b,d) optical conductivity for ${\cal C}=1$ and ${\cal C}=0$, respectively. In this figure, the dashed vertical line marks the position the TSS gap. At photon energies much smaller than the TSS gap value ($\hbar \omega << \textbf{$E_{\rm gap}$}$), we find $\Re\sigma_{xy}$ to be quantized in units of $e^2/h$ (Hall plateau) which is a characteristic of the Chern insulating (${\cal C}=1$) state (see Fig. \ref{fig:fig4}a). On the other hand, $\Im\sigma_{xy}$ remains zero throughout the gapped region. As the frequency increases and approaches the threshold of $\hbar \omega \approx \textbf{$E_{\rm gap}$}$, photons have enough energy to excite electrons across the gap. In the region of $\hbar\omega> E_{\rm gap} $, $\Re\sigma_{xy}$ increases beyond the quantized value, and $\Im\sigma_{xy}$ becomes greater than zero. As shown in Fig.~\ref{fig:fig4}b for the ${\cal C}=1$ case, the real part of the longitudinal conductivity ($\Re\sigma_{xx}$) remains zero inside the gapped region due to the absence of interband transitions, signaling absence of dissipation, whereas in the the same region $\Im\sigma_{xx}$ decreases linearly. These findings are in qualitative 
agreement with the results of Ref.~\cite{lei2023kerr} based on a simplified coupled Dirac-cone model. 

It is pertinent to mention that a non-zero Berry curvature acts as an effective magnetic field in momentum space, generating an anomalous transverse velocity and thus a Hall response in the gapped TSS. The quantization of $\Re\sigma_{xy}$ in the parallel spin alignment configuration of the top and bottom SL layers, where each surface contributes to half-quantized Hall conductance, is a signature of the QAHE in the zero frequency limit. 
In the  $C=0$ state, i.e. when the surface SL spins are aligned in the anti-parallel direction, both $\Re\sigma_{xy}$ and $\Im\sigma_{xy}$ are zero in the TSS gap region leading to the so-called zero-Hall plateau of the zero-frequency Hall conductance in the axion insulating phase \cite{mogi2017tailoring}.
Alternatively, a non-zero $\Re\sigma_{xy}$ reflects the asymmetric optical response of the material to right- and left-circularly polarized light in systems where time-reversal symmetry is broken, thereby producing a rotation of the plane of light polarization.

Since MBT thin films intrinsically break time-reversal symmetry through their internal magnetic order, magneto-optical (MO) measurements provide an indirect probe to confirm these transport features. Indeed, the coupling between circularly polarized incident light and spin-flip states should lead to a distinct MO response in the two phases, in the form of Faraday rotation ($\theta_F$) and Kerr rotation ($\theta_K$) angles, corresponding to the rotation of the polarization plane of transmitted and reflected light from MBT samples. This behavior can be described using a simple electrodynamic model. If a thin film of thickness $d$ ($d\ll\lambda$, with $\lambda$ being the wavelength of incident light) is placed between two dielectrics with refractive indices $n_1=\sqrt{\varepsilon_1}$ (top) and $n_2=\sqrt{\varepsilon_2}$ (bottom), the MO response is characterized by two-dimensional conductive layers with surface conductivity $\sigma$ (measured in units of $e^2/h$). Imposing the boundary conditions allows us to determine the transmission field ($E_{x/y}^{t}$) and reflection field ($E_{x/y}^{r}$). 

We then express these fields in a linear polarization basis:
\begin{align}
\binom{E_x^{t}}{E_y^{t}}
&= \frac{1}{N}
\binom{\,2 n_1\!\left(n_1+n_2+2\alpha\,\sigma_{xx}\right)\,}{-\,4\alpha n_1\,\sigma_{xy}},\label{eq-2}\\[4pt]
\binom{E_x^{r}}{E_y^{r}}
&= \frac{1}{N}
\binom{\,n_1^{2}-\!\left(n_2+2\alpha\,\sigma_{xx}\right)^{2}-\left(2\alpha\,\sigma_{xy}\right)^{2}\,}{-\,4\alpha n_1\,\sigma_{xy}},\label{eq-3}
\end{align}
where,
$N=\left(n_1+n_2+2\alpha\,\sigma_{xx}\right)^{2}+\left(2\alpha\,\sigma_{xy}\right)^{2}$.

Using Eq.~\eqref{eq-2} and \eqref{eq-3} together with the optical conductivity tensor computed via the Kubo-Greenwood formula of Eq.~\eqref{eq-1}, we finally obtain expressions for the angles $\theta_F$ and $\theta_K$ as follows:
\begin{equation}
\theta_F=\tfrac{1}{2}\!\left(\arg E_{+}^{t}-\arg E_{-}^{t}\right),\ 
\theta_K=\tfrac{1}{2}\!\left(\arg E_{+}^{r}-\arg E_{-}^{r}\right),\label{eq-4}
\end{equation}
where $E_{\pm}^{\,t,r}=E_x^{\,t,r}\pm i\,E_y^{\,t,r}$.

\begin{figure}[!b]
    \centering
    \includegraphics[width=8.5cm]{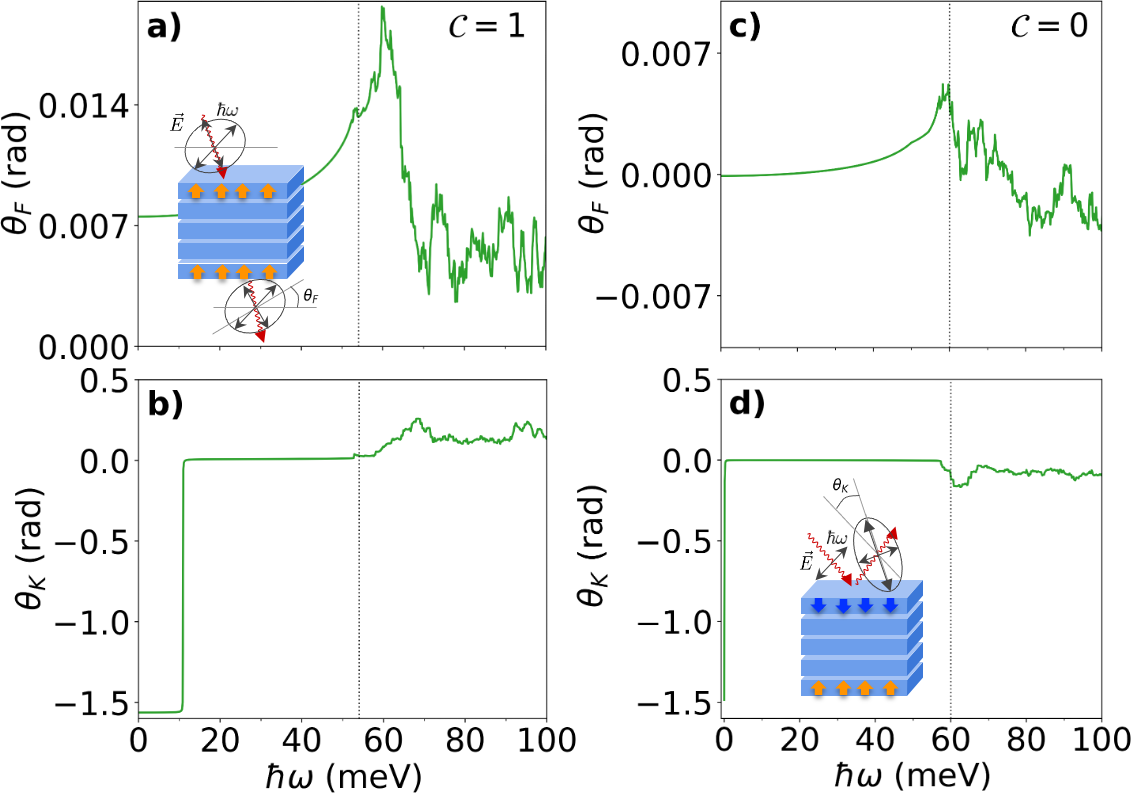}
    \caption{Magneto-optical response of two representative spin-flipped 5-SL MnBi$_2$Te$_4$ cases. (a,c) Faraday rotation angle, and (b,d) Kerr rotation angle (in radian (rad)) against optical frequency ($\omega$) for the representative ${\cal C}=+1$ and ${\cal C}=0$ cases.}
    \label{fig:fig5}
\end{figure}

We use these expressions to 
calculate the MO response of spin flip states, focusing on the $\theta_F$ and $\theta_K$ angles for ${\cal C}= +1$ and ${\cal C}=0$ cases, as shown in Fig.~\ref{fig:fig5}. In the limit $\hbar \omega << \textbf{$E_{\rm gap}$}$ (dotted black line), $\theta_F$ acquires the ``quantized" value of $\theta_F=\cal C \;$tan$^{-1}\alpha$ ($\alpha$ is the fine structure constant) for ${\cal C}= +1$ case, whereas it vanishes for ${\cal C}= 0$ case, as shown in Fig.~\ref{fig:fig5}(a) and (c). These observations can be explained by the specifics of the measurements to underlying physical processes, as depicted schematically in the inset of Fig.~\ref{fig:fig5}a. For the ${\cal C}= +1$ case, the spins on the top and bottom SLs are parallel. Since $\theta_F$ is proportional to the sum of the Hall conductivities along the light propagation direction, the contributions of both surfaces to the Faraday angle add up, resulting in a finite rotation of the polarization plane. Conversely, for the ${\cal C}= 0$, the spins on the top and bottom surfaces are oppositely oriented. As a consequence, the polarization-plane rotation induced at the top surface is compensated by an opposite rotation at the bottom surface, leading to a vanishing net $\theta_F$ in the gapped region, as shown in Fig.~\ref{fig:fig5}c.

On the other hand, light reflection from MBT spin-flip states can be studied using the Kerr angle dynamics, as displayed in Fig.~\ref{fig:fig5}(b) and (d). 
For spin flip states leading to ${\cal C}= 0 $ (panels with red color in Fig. \ref{fig:fig3}(a)), $\theta_K$ is identically zero as shown in Fig.~\ref{fig:fig5}(d), just like the even-numbered MBT films studied in Ref.~\cite{lei2023kerr}. Alongside, Fig.~\ref{fig:fig5}(b) shows that systems with spin flip states leading to ${\cal C}= 1$ (panels with blue color in Fig. \ref{fig:fig3}(a)) are characterized by non-zero value of $\theta_K$. In the DC limit ($\omega \to 0$), neglecting dissipation terms ($\Im\sigma_{xy}$ and $\Re\sigma_{xx}$), we can estimate $\theta_K\simeq -\pi/2$, in agreement with the results for odd-numbered MBT films in  Ref.~\cite{lei2023kerr}. Notice that the value of $\theta_K$ remains constant - generating the flat plateau visible in Fig.~\ref{fig:fig5}(b), but drops sharply to zero at a finite frequency. Importantly, the $\theta_K$ sudden jump to zero obtained from the realistic TB model does not correlate to the TSS energy gap but occurs at a considerably smaller threshold of $\hbar\,\omega \approx 10$ meV. This behavior of the $\theta_K$ differs from the results of Ref. \cite{lei2023kerr} where the size of the flat plateau is controlled by TSS energy gap. In other words, in the description of the Kerr dynamics ($\theta_K (\omega)$) we find a discrepancy between the microscopic TB model used here and the coupled Dirac-cone model, although both models capture correctly the topological phase of the system. The reason of this discrepancy is explained below.

We first address the sharp change from $-\pi/2$ to 0 of $\theta_K$ (presented in Fig.~\ref{fig:fig5}b) and the decrease in the width of the plateau below the TSS gap found in the microscopic TB model. For this, it is necessary to start from Eq.~\eqref{eq-4} that gives the Kerr angle $\theta_K$ in as the difference of $\arg z=(\arg E_{+}^{r}-\arg E_{-}^{r})$. Here $\arg z$ is the phase of the complex number $z$ in radians in the interval $(-\pi, \pi)$. The phase of a complex number $z$ can in principle be expressed as the inverse tangent of the ratio $(\Im z/\Re z)$. Note however that a naive calculation of arctan$(\Im z/\Re z)$ would only yield an angle in the interval $(-\pi/2, \pi/2)$. In order to get the full range of the phase, it is necessary to take into account the signs of $(\Re z)$ and $(\Im z)$ separately, which allows us to determine the quadrant on which the complex number resides. As shown in detail in the Supplementary Information (see Eq. S4-S7), this is conveniently done by means of the two-argument function \textbf{atan2}$(\Im z,\Re z)$\cite{organick1963fortran}. This algorithm uniquely specifies the argument of $z$ in the interval $-\pi$ to $\pi$ and thereby correctly reproduces branch switches caused by the sign changes of $(\Re z)$ and $(\Im z)$, which are ultimately responsible for the sharp change in $\theta_K$.

By writing $\theta_K$ in terms of the \textbf{atan2} function, we finally obtain the following expression:

\begin{equation}
\theta_K = \frac{1}{2}\textbf{atan2}\big[\Im(E_+^r(E_-^r)^*),\Re(E_+^r(E_-^r)^*)\big]
\label{eq-5}
\end{equation}
Calculating $\Im(E_+^r(E_-^r)^*)$ and $\Re(E_+^r(E_-^r)^*)$ in the TSS-gapped region
($\Re\sigma_{xx}=0$ and $\Im\sigma_{xy}=0$) and using the piecewise definition of the \textbf{atan2} function
(see Supplementary Eqs. S9-S10), one can directly track how the conductivity components control the behavior of $\theta_K$.
Inside the gap, $\Re\sigma_{xy}$ remains constant, whereas $|\Im\sigma_{xx}|$ increases monotonically with $\omega$.
As a result, the complex quantity $z(\omega)=E_+^r(E_-^r)^* (\omega)$ evolves continuously in the complex plane, and the Kerr angle
$\theta_K(\omega)=\frac12\,\arg z(\omega)$ decreases toward zero as the relative phase between the reflected circular components
$E_+^r$ and $E_-^r$ is reduced by the growing longitudinal reactive response. The sharp step-like crossover toward $\theta_K\simeq 0$ occurs at the point where the reactive longitudinal contribution becomes
comparable to the Hall term, causing $\Re z(\omega)=0$ to vanish and change sign. This produces a rapid phase shift of the corresponding reflected circular component, i.e., a rapid change of the relative phase
$\arg\!\big[E_+^r(E_-^r)^*\big]$ by approximately $\pm\pi$, which translates into a change of $\theta_K$ by $\pm\pi/2$.
In the $\mathrm{atan2}$ representation, this phase shift appears as the branch switching discussed in  Figs. S2-S3 of the Supplementary.
The apparent abruptness of the step is set by how quickly this compensation is reached, which we discuss in more detail in the next section.

\subsection{Comparison with the coupled Dirac-cone model}

To interpret the characteristic features of the MO response in different regimes, we use a coupled Dirac-cone model introduced in Ref.~\cite{lei2020magnetized} and based on the theoretical formalism developed earlier in Ref.~\cite{tse2011magneto}. Therein, the properties of MBT thin films are modeled by coupled Dirac cones originating from the top and bottom SLs, and are described by the following Hamiltonian:
\begin{align}
&H = \sum_{\mathbf{k}_\perp,i,j}
\Big[
\big( (-1)^{i}\,\hbar v_D \,(\hat{\mathbf z}\!\times\!\boldsymbol{\sigma})\!\cdot\!\mathbf{k}_\perp
+ m_i\,\sigma_z \big)\,\delta_{ij}
\;+\nonumber\\
&\; \quad\quad\quad\quad\quad\Delta_{ij}\,(1-\delta_{ij})
\Big]\,c^\dagger_{\mathbf{k}_\perp i}\,c_{\mathbf{k}_\perp j}
\end{align}

\begin{figure*}[!]
    \centering
    \includegraphics[width=18cm]{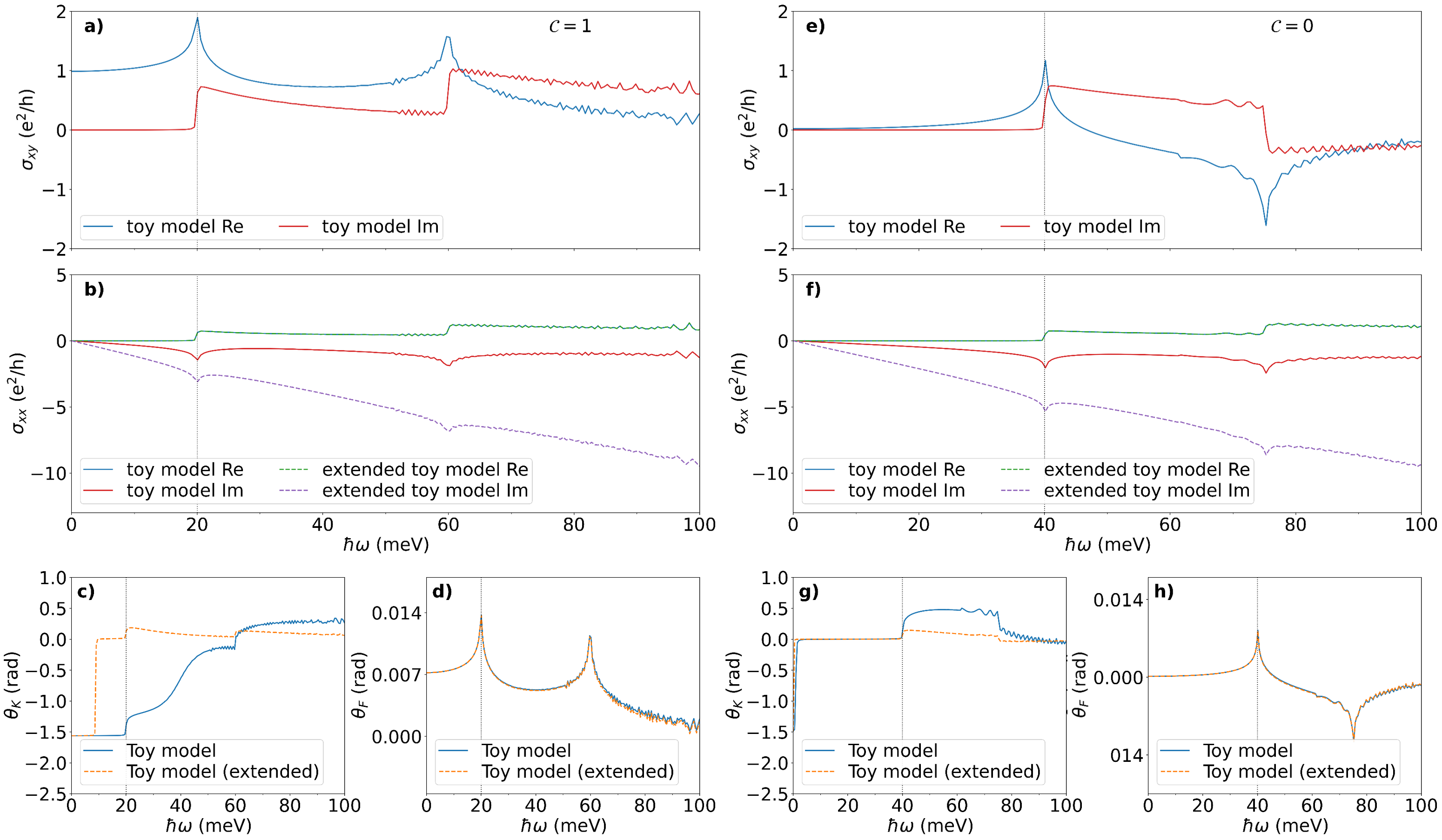}
    \caption{Frequency-dependent optical and magneto-optical response of the coupled Dirac cone model for our representative (a-d) ${\cal C}=+1$ and (e-h) ${\cal C}=0$ cases. (a, e) Transverse (Hall) conductivity ($\sigma_{xy}$) and (b, f) Longitudinal conductivity ($\sigma_{xx}$) (in units of $e^2/h$); real and imaginary parts are distinguished by different colors as indicated in the legends. Solid lines correspond to the coupled Dirac cone model (toy model), while dashed lines show the extended model including a phenomenological reactive correction $-i\kappa\omega$. (c, g) Kerr rotation $\theta_K(\omega)$ and (d, h) Faraday rotation $\theta_F(\omega)$ angles obtained from ($\sigma_{xy}$) and ($\sigma_{xx}$). Vertical dotted lines mark the TSS energy gap value.
}
    \label{fig:fig6}
\end{figure*}

Here, $\mathbf{k}_\perp$ is the in-plane two-dimensional quasi-momentum, $c^\dagger_{\mathbf{k}_\perp i}$ and $c_{\mathbf{k}_\perp j}$ are the electron creation and annihilation operators in the i-th and j-th Dirac cones, respectively. Indexes i and j are numbered for each SL, odd index corresponds to the upper surface and the even index to the lower surface. The Dirac kinetic term $\hbar v_D \,(\hat{\mathbf z}\!\times\!\boldsymbol{\sigma})\!\cdot\!\mathbf{k}_\perp$ describes the linear dispersion of surface states and the factor $(-1)^{i}$ takes into account the opposite sign of the Dirac term for the upper and lower surfaces. The mass term $m_i\,\sigma_z$ determines the exchange splitting of the corresponding Dirac cone, where $m_i$ is formed as the sum of contributions from  nearby Mn magnetic layers within the same SL ($J_S$ parameter) and from neighboring SL ($J_D$ parameter). The off-diagonal elements $\Delta_{ij}$ describe the hybridization between different Dirac cones. Specifically, we use $\Delta_{S}$ to indicate the coupling between upper and lower cones within a SL and a $\Delta_{D}$ the  coupling between cones across the gap between adjacent SL. Thus, a N-layer film is reduced to a quasi-2D system with 4N bands. The following set of parameters are used in our model calculations: $v_D=5\times10^5$ m/s, $\Delta_{S} =84$ meV, $\Delta_{D} = -127$ meV, $J_S=36$ meV, $J_D=29$ meV. We employed the Kubo-Greenwood formula (Eq. \eqref {eq-1}) to obtain the components of  conductivity tensor. Finally, Eq.~\eqref{eq-2}-\eqref{eq-5}, used above for the DFT TB model, are also used here to calculate the Kerr and Faraday rotation angles.

The conductivity plots for this model are shown in Fig.~\ref{fig:fig6}(a,b,e,f). We notice that the coupled Dirac-cone model captures qualitatively the behavior of the TB model, with two main differences:

i. The main peaks appear exactly at the threshold set by energy gap TSS $E_{\rm gap}$, whereas in the TB model they appear at frequencies exceeding $E_{\rm gap}/\hbar$; 

ii. The imaginary part of the conductivity $\Im\sigma_{xx}$ is approximately constant and small inside the TSS gap region (Fig.~\ref{fig:fig6}(b,f), red line) before developing a sharp cusp at $\omega = E_{\rm gap}/\hbar$, in contrast to the linear decrease with a relative large negative slope displayed by the TB model (see Fig.~\ref{fig:fig4}(b,d)). This is observed to occur for both  ${\cal C}=1$ and ${\cal C}=0$. It turns out that this qualitative and quantitative difference between the two models is responsible for the remarkably different behavior of their corresponding Kerr angles, for the ${\cal C} = 1$, displayed in Fig.~\ref{fig:fig6}(b) and Fig.~\ref{fig:fig6}(c)(blue line), respectively. For both models $\theta_K$ starts at $-\pi/2$ for  $\omega = 0$ forming a flat plateau, whose width, however, is different for two cases. Using the coupled Dirac cone model, the threshold is exactly equal to the TSS gap, while for the TB model the value is considerable smaller than the gap. Beyond this threshold, $\theta_K$ for the TB model is just a step function sharply dropping to zero and remaining so until the $\hbar \omega \approx $ TSS gap. For the coupled Dirac-cone model, $\theta_K$ smoothly tends to zero after a small sharp step up.

Point (ii) is essentially the result of the more complex electronic structure of the microscopic TB model with respect to the one of the coupled Dirac-cone model. Indeed, the band structure of the former, shown in Fig.~\ref{fig:fig2}, contains many more bands than the latter, which for a 5-SL MBT film consists of only 20 bands. (Remember that in this model a N-SL thin film is reduced to a quasi 2D-system with 4N bands).
Indeed, the considerably more complex band structure of the TB model allows many more transitions when calculating $\Im\sigma_{xx}$ (the reactive response term of $\sigma_{xx}$ given by the principal value), leading to a linear behavior with a large slope within the TSS gap region  \footnote{This can be understood by looking at the analytical expressions of the conductivity tensor derived from the coupled Dirac-cone model, see (See Eqs.~S10$-$S13 in the Supplementary Information). Indeed,
both $\Im\sigma_{xx}(\omega) $ and $\Re\sigma_{xy}(\omega)$ contain logarithmic and inverse tangential terms, generating sharp peaks and steps. Furthermore, the mass term ($m$) in these expressions defines the depth of the peak, followed by a smooth logarithmic tail. The logarithmic function itself tends to zero, but as the number of transitions increases, the logarithmic tail is truncated. Each transition introduces a more negative value into $\Im\sigma_{xx}$, leading to an overall linear decrease of the function, the slope of which becomes steeper by adding transitions.}.

\begin{figure}[!]
    \centering
    \includegraphics[width=8.5cm]{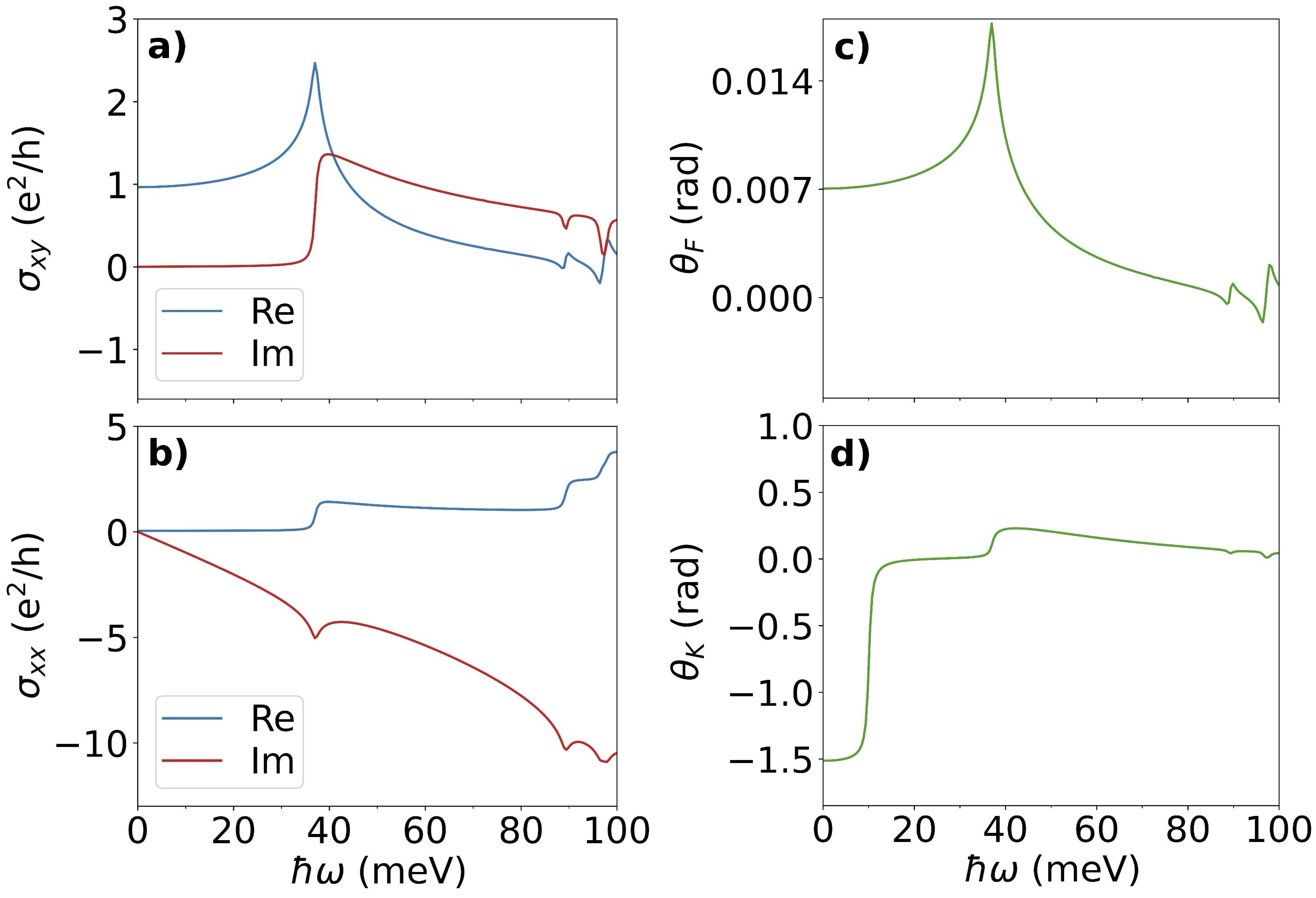}
    \caption{Optical conductivity and MO response of the coupled Dirac-cone model for an increased number of layers (N=31). (a) Transverse (Hall) conductivity ($\sigma_{xy}$), (b) Longitudinal conductivity ($\sigma_{xx}$) (in the units of $e^2/h$). Real and imaginary parts are shown in blue and red colors, respectively. (c) Faraday rotation $\theta_F(\omega)$ and (d) Kerr rotation $\theta_K(\omega)$ angles.}
    \label{fig:fig7}
\end{figure}

To demonstrate this point, we plot the conductivity tensor components ($\sigma_{xx}$ and $\sigma_{xy}$) together with the corresponding Kerr and Faraday rotation angles in Fig.~\ref{fig:fig7}, calculated for a 31-SL coupled Dirac-cone model and whose electronic structure now consists of 124 energy bands that account for an increased number of transitions. We can see that the transverse component ($\sigma_{xy}$) shows good agreement with the results of the TB model (see Fig.~\ref{fig:fig4}(a)). Hence, $\theta_F$ remains unchanged and retains its quantized value as shown in  Fig.~\ref{fig:fig7}(c). In addition to it, the imaginary part of the longitudinal conductivity ($\Im\sigma_{xx}$) now exhibits a steeper slope in contrast to the model calculation shown in red color in Fig.~\ref{fig:fig6}(b). 

A steeper (more negative) slope of $\Im\sigma_{xx}(\omega)$ implies a faster growth of the reactive longitudinal response inside the gap. This accelerates the evolution of the complex quantity $z(\omega)=E_+^r(E_-^r)^*$ in the complex plane and brings it to the
branch-switching condition $\Re z(\omega)=0$ at a lower frequency (see Supplemental Eq. S10 and Figs. S2-S3). This corresponds to reaching the point where the longitudinal reactive contribution becomes comparable to the Hall term,
so that the relative phase between the reflected circular components changes rapidly. As a result, the value of the phase extracted
via \textbf{atan2} belonging to the first branch switches branch earlier, and $\theta_K(\omega)$ exhibits an earlier and sharper step-like crossover toward values close to
zero (see Supplementary Fig.~S3(b--d)). This reduces the width of the ($\theta_K = -\pi/2$)-plateau
emanating from $\omega = 0$
and accounts for the step-like feature observed in the microscopic TB results. 

To elucidate further these results, we introduce an extended coupled Dirac-cone model in which a phenomenological, \textit{ad-hoc} linear imaginary term is included in the longitudinal component of the conductivity ($\sigma_{xx}$), in order to capture the observed frequency dependence: 
\begin{equation}
\sigma^{\rm extd}_{xx}(\omega) =  \sigma_{xx}(\omega)  - i\kappa\omega
\end{equation}
where $\kappa$ is a phenomenological parameter. This modified expression mimics the linear dependence of $\Im \; \sigma_{xx}$ found in the TB model. 
As a result, it can be seen that for the conductivity component $\Im\sigma_{xx}^{\rm extd}$, the slope of the graph increases significantly ((Fig.~\ref{fig:fig6}(b, f)(purple line). Using this extended model, the Kerr and Faraday angles were obtained for both cases ${\cal C}=0$ (Fig. \ref{fig:fig6}(g,h)) and ${\cal C}=1$ (Fig. \ref{fig:fig6}(c,d)). In this case, we can see that the plateau's width exactly matches the value of the energy gap, and the switch to zero is smoother than in the TB model. 

This difference is directly related to how fast the complex quantity $z(\omega)=E_+^r(E_-^r)^*$ reaches the branch-switching condition
$\Re z(\omega)=0$ (see Supplementary Figs. S2-S3). For finite $\kappa$, the crossing occurs within a narrow frequency interval and the
resulting step in $\theta_K$ appears almost discontinuous. In contrast, for $\kappa=0$ the trajectory may first approach the real axis
(i.e., $|\Im z(\omega)|$ becomes small), leading to a smooth decay of $\theta_K$ before the final rapid collapse when the sign change of
$\Re z(\omega)$ is completed. Accordingly, increasing $\kappa$ sharpens the crossover and reproduces the more abrupt step-like behavior
seen in the TB results by effectively accelerating the evolution of $\Im\sigma_{xx}$.

In contrast, our linear addition eliminates this difference by modifying the $\Im\sigma_{xx}$ component, resulting in the behavior of the extended model matching that of the TB model. For the case ${\cal C}=0$, both models show good agreement with the TB model. In real materials, the conductivity can be influenced by bulk charge carriers, an electric field, a substrate, and a dielectric background. All these effects act in a similar way by increasing in the number of interband transitions. Our additional term in the model reflects the functional trend of such effects.\\

\section{Conclusions}

In summary, we have studied the topology and magneto-optical (MO) response of various spin-flip states of a 5-SL MBT thin film by means of TB models obtained from first-principles calculations and a coupled Dirac-cone model. We have shown that a non-zero net magnetization ($\sim 5\,\mu_B$) does not guarantee the topological phase of 5-SL MBT. Instead, depending on the relative spin-orientation of the top and bottom SLs, a 5-SL can realize both a Chern insulating phase with ${\cal C}=+1$ and ${\cal C}=0$. The magnitude of the TSS gap is likewise controlled primarily by the local surface spin alignments rather than by a global magnetic order. We further demonstrated that these distant topological phases are directly reflected in the MO response. In the low frequency limit and within the TSS gap, the Faraday rotation exhibits a quantized response for ${\cal C}=+1$ phase and vanishes for ${\cal C}=0$, providing a robust probe of topology. In contrast, the Kerr rotation is found to be highly sensitive to the detailed frequency dependence of the longitudinal optical conductivity ($\sigma_{xx}$). By comparing realistic TB results with a coupled Dirac-cone model, we clarified the limitations of the latter in describing Kerr rotation dynamics in spin-flip states and discuss the underlying mechanism responsible for rapid collapse of the Kerr plateau. Our results highlight the crucial role of surface-resolved magnetism in magnetic TIs and provide guidance for interpreting MO measurements on MBT thin films under different spin-flip configurations.

\section{Acknowledgment}
The work is financially supported by the
Swedish Research Council (grant no: VR 2021-04622).
The computations were enabled by
resources provided by the National Academic Infrastructure for Supercomputing in Sweden (NAISS) partially
funded by the Swedish Research Council through grant
agreement no. 2022-06725 and the Centre for Scientific and Technical Computing at Lund University (LUNARC).
Work in Austin was supported by the W. M. Keck Foundation under grant 996588.

\bibliography{main.bib}
\end{document}